\documentclass{astlb} 

\usepackage[hyperfootnotes=false]{hyperref}
\usepackage[utf8]{inputenc}
\usepackage[T2A]{fontenc}
\usepackage{color}
\usepackage{epsf}
\usepackage{amsmath}
\usepackage{xcolor}
\usepackage{soul}

\usepackage{amsfonts}
\usepackage{amssymb}
\usepackage{epsf}
\usepackage{graphicx}
\usepackage{gensymb}
\usepackage{float}
\usepackage[normalem]{ulem}
\usepackage{enumitem}
\usepackage{pdflscape}

\def\srg{\textit{SRG}}
\def\art{\emph{ART-XC}}
\def\ero{\emph{eROSITA}}

\def\flx{erg\,cm$^{-2}$\, s$^{-1}$}

\begin{document}

\journalinfo{2025}{00}{0}{0}[0]
\UDK{524.77}

\title{SRG/ART-XC All-Sky X-ray Survey: Sensitivity Assessment Based on Aperture Photometry}

\author{
  N.~Y.~Tyrin\address{1},
  R.~A.~Krivonos\email{krivonos@cosmos.ru}\address{2},
  V.~A.~Arefiev\address{2}, 
  R.~A.~Burenin\address{2},  
  E.~I.~Zakharov\address{2},  
  A.~A.~Lutovinov\address{2}, 
  S.~Y.~Sazonov\address{2},  
  A.~D.~Samorodova\address{2}, 
  E.~V.~Filippova\address{2}, \\
  A.~A.~Abbakumov\address{2}, 
  V.~V.~Konoplev\address{2},
  F.~V.~Korotkov\address{2},
  V.~N.~Nazarov\address{2}
    \addresstext{1}{Kazan Federal University, Kazan}    \addresstext{2}{Space Research Institute of the Russian Academy of Sciences, Moscow}
}

\shortauthor{Tyrin et al.}

\shorttitle{SRG/ART-XC all-sky survey upper limits}

\submitted{00.00.2025 г.}

\begin{abstract}

The Spectrum-Roentgen-Gamma (SRG) observatory continues to operate successfully in orbit at the Lagrange point L2. The {\it Mikhail Pavlinsky} {\art} telescope has demonstrated high efficiency in conducting X-ray surveys both over large sky regions and the entire celestial sphere. A recently published source catalog, based on data from the first four and partially completed fifth sky scans, contains 1,545 objects detected in the 4–12 keV energy range. In this work, using the same sky survey data, we assess the sensitivity to point source detection across the celestial sphere based on X-ray aperture photometry — that is, we calculate the upper flux limit in the 4–12 keV band at any given significance level. The method is implemented using both Poisson statistics and Bayesian inference, with consistent results between the two approaches. This information is important for studying variable and transient X-ray sources, as well as sources that are not detected with sufficient statistical significance in the {\art} all-sky survey. The {\art} upper limit service is available at  \url{https://www.srg.cosmos.ru/uplim}.

  \keywords{sky surveys, X-ray sources} 

\end{abstract}

\section*{Introduction}

\noindent

The Spectrum-Roentgen-Gamma \citep[SRG,][]{srg21} space observatory was designed to conduct wide angle X-ray surveys with better sensitivity, angular resolution and energy characteristics than its' predecessors, such as Uhuru (1970-1973, \cite{forman1978}), HEAO-1 (1977-1979, \cite{1984ApJS...54..581L}) and ROSAT (1990-1991, \cite{rosat}. 
The observatory was launched with Proton rocket on July 13, 2019, from the Baikonur Cosmodrome to a halo orbit near the Sun-Earth L2 Lagrange point. Scientific observations from this vantage point commenced on December 12, 2019.
The spacecraft carries two X-ray telescopes: {\ero} \citep{erosita} and the {\it Mikhail Pavlinsky} {\art} \citep{artxc}.
Each telescope comprises an assembly of seven coaxial, independent modules containing X-ray mirrors and detectors, enabling a significant increase in the effective collecting area. {\ero} and {\art} complement each other by providing sensitivity in the energy ranges of 0.2–8 keV and 4–30 keV, respectively.

Since commencing operation in July 2019, {\art} has shown stable performance consistent with pre-flight expectations.
Several X-ray surveys were conducted during the telescope’s calibration and performance verification phase: a survey of the Galactic plane near Galactic longitude $l\simeq20^{\circ}$ \citep{2023AstL...49..662K} and a deep survey of the Galactic bulge \citep{2024MNRAS.529..941S}.
Based on the first two all-sky surveys conducted over the time period between December 2019 and December 2020, a catalog of sources discovered by {\art} in the 4-12 keV energy range was constructed \citep[ARTSS1-2,][]{2022A&A...661A..38P}.
This catalog was then updated using data spanning the period from December 2019 to March 2022, corresponding to ${\sim}4.4$ all-sky surveys \citep[ARTSS1-5,][]{2024A&A...687A.183S}.
It now contains 1545 X-ray sources.

{\art} all-sky X-ray surveys establish a valuable legacy for the SRG observatory and make a significant contribution to X-ray astronomy.
However, many true astrophysical X-ray sources remain hidden within the observational noise and fail to reach the detection threshold set for the catalogs.
Such missed X-ray sources may simply be faint or be variable in nature.
This is why being able to calculate an upper limit on the source flux for arbitrary celestial sphere coordinates would be a valuable addition to the source catalog.
This information would also be valuable for studying sources initially discovered in other ranges of the electromagnetic spectrum.
Consequently, upper limits derived from the {\art} all-sky survey data may be valuable for studying long-period X-ray variability, conducting transient searches, and multiwavelength studies of sources and source populations.

This study presents a methodology for calculating upper limits and measuring X-ray flux using aperture photometry with data from the {\art} telescope in the 4-12 keV range, based on four all-sky surveys (ARTSS1-4).
A comparable study, using {\ero} data from the first all-sky survey (December 2019 - June 2020) by the SRG observatory, for the hemisphere $180^{\circ} \leq l \leq 360^{\circ}$, is presented in \cite{2024A&A...682A..35T}.


\section*{Methods}


\noindent
To determine the point source flux confidence limits we used aperture photometry.
We extracted source counts from an aperture of radius $71''$ centered on the requested coordinates. This corresponds to the W90 radius -- the radius containing 90\% of the total flux -- of the {\art} telescope's point spread function (PSF) in survey mode \citep{2025arXiv250513296K}.
The number of counts detected in such an aperture is described by a Poisson distribution
\begin{equation}
    P(N|S+B) = \frac{(S+B)^N\cdot e^{-(S+B)}}{N!},
\end{equation}
where $P$ is the probability of observing $N$ counts, with the expected number of counts in the aperture being $S+B$.
Here, $S$ represents the number of counts attributed to the source, and $B$ represents the background.
For the background region, we selected an annulus centered on the requested coordinates, with the inner and outer radii of $213"$ and $355"$ (3 and 5 source aperture radii, respectively).
We estimated the expected number of background counts, $B$, by scaling the average count rate in the background region by the area of the source aperture.
In the following statistical derivations, the uncertainty in the background count rate was assumed to be negligible.

To estimate upper and lower flux limits, one can employ either the classical frequentist approach \citep{1986ApJ...303..336G} or a Bayesian method \citep{kraft1991}.
In contemporary literature, the latter is typically preferred \citep{2022MNRAS.511.4265R,2024A&A...682A..35T}. 
However, we implemented both methods to ensure transparency and allow for cross-checking of the results.

To compute the classical frequentist confidence limits, the equations from \cite{1986ApJ...303..336G} were used.
One-sided upper ($\lambda_u$) and lower ($\lambda_\ell$) limits for $N$ detected counts are defined as follows:
\begin{equation}
    \sum_{x=0}^N \frac{\lambda_u^x e^{-\lambda_u}}{x!} = 1 - CL, \label{gehrelssum1}
\end{equation}
\begin{equation}
    \sum_{x=0}^{N-1}\frac{\lambda_\ell^x e^{-\lambda_\ell}}{x!} = CL, \label{gehrelssum2}
\end{equation}
\noindent
where $CL$ is the required confidence level.
A double-sided confidence interval with confidence level $CL$ can be constructed by replacing $CL$ with $CL_{2S}=(1+CL)/2$ in these expressions.

When using the classical approach, to account for background one should subtract the background counts $B$ from the upper and lower limits after computing them \citep{kraft1991}.
Then, the double-sided confidence limit (from $S_\ell$ to $S_u$) and the one-sided upper limit ($S_U$) for source counts $S$, background counts $B$, and confidence level $CL$ are as follows:

\begin{equation}
    S_u^c = \lambda_u(CL_{2S})-B,
\end{equation}
\begin{equation}
    S_\ell^c = \lambda_\ell(CL_{2S})-B,
\end{equation}
\begin{equation}
    S_U^c = \lambda_u(CL)-B.
\end{equation}

Calculation of Bayesian confidence intervals was implemented using expressions from \cite{kraft1991}.
This method leverages Bayes' theorem,
\begin{equation}
    f_{N,B} \propto p(S)P_S(N)
\end{equation}
\citep{10.1063/1.3127954}. 
Here $f_{N,B}$ represents the posterior probability, which incorporates prior information about the distribution, contained in $p(S)$, i.e., the prior. 
To maintain objectivity, the prior in \cite{kraft1991} only constrains the flux to be non-negative.
In this case, substituting the Poisson distribution and a constant prior into Bayes' theorem yields
\begin{equation}
    f_{N,B}(S) = C \frac{e^{-(S+B)}(S+B)^N}{N!}
\end{equation}
for $S\ge0$. 
The normalization constant $C$ is defined as
\begin{equation}
    \begin{aligned}
        C &= \bigg[\int_0^\infty \frac{e^{-(S+B)}(S+B)^N}{N!}dS\bigg]^{-1}= \\
        & =\sum_{n=0}^{N}(\frac{e^{-B}B^n}{n!})^{-1}.
    \end{aligned}
\end{equation}
Confidence limits can then be obtained by integrating the posterior probability $f_{N,B}$ over $S$ and finding such $S_\ell^b$ and $S_u^b$ that:
\begin{equation}
    \int_{S_\ell^b}^{S_u^b} f_{N,B}(S)dS=CL.
\end{equation}
\noindent
In the special case of a one-sided upper limit:
\begin{equation}
    \int_{0}^{S_U^b} f_{N,B}(S)dS=CL.
\end{equation}

For a two-sided confidence interval, the choice of limits is somewhat arbitrary.
We chose the interval's length being minimal as the criteria for its construction. 
In this case, it implies that the probability density is equal at both ends of the interval \citep{kraft1991}:
\begin{equation}
    f_{N,B}(S_\ell^b)=f_{N,B}(S_u^b).
\end{equation}

The limits derived using both methods were then converted to count rate $r$ by dividing the counts by the average exposure within the aperture, $t$, and the enclosed energy fraction, $EEF$, which is equal to $0.96$ for a $71"$ aperture:
\begin{equation}
    r = \frac{S}{t\cdot EEF}
\end{equation}

The X-ray flux was then calculated by multiplying the count rate by the conversion factor $CF$, equal to $4\cdot10^{-11}$\flx \citep{2024A&A...687A.183S}:
\begin{equation}
    F = r \cdot CF.
\end{equation}

\section*{Implementation}
\noindent
However, since the expressions for $S_\ell$, $S_u$ and $S_U$ cannot be analytically obtained using the above equations, numeric methods are required.
Equations \eqref{gehrelssum1} and \eqref{gehrelssum2} can be expressed in terms of the regularized upper incomplete gamma function $Q$ as follows:
\begin{equation}
    Q(N+1,\lambda_u)=1-CL,
\end{equation}
\begin{equation}
    Q(N,\lambda_\ell)=CL.
\end{equation}
We used the implementation of the inverse of $Q$ available in the \texttt{scipy} package \citep{2020NatMe..17..261V} to compute $\lambda_U$, $\lambda_u$ and $\lambda_\ell$.

To compute Bayesian confidence limits, we used the method outlined in \cite{kraft1991}.
Two-sided intervals were computed via the \texttt{poisson\_conf\_interval} function from the \texttt{scipy} package \citep{Astropy2022ApJ...935..167A}.
For the one-sided upper limit, we used the following expression:
\begin{equation}
\begin{aligned}
    S_U^b &= P^{-1}\Big[N+1,CL\cdot Q(N+1, B)+\\
    &+P(N+1,B)\Big],
\end{aligned}
\end{equation}
where $P$ is the regularized lower incomplete gamma function, for which we also utilized the implementation from the \texttt{scipy} package.

The flux was estimated as follows:
\begin{equation}
    F = \frac{N-B}{t\cdot EEF}\cdot CF,
\end{equation}
which, under the adopted assumptions (neglecting background uncertainty and assuming a constant prior distribution), corresponds to the most likely flux value in both frequentist and Bayesian statistics.
It was found that this estimate of aperture fluxes for the positions of known sources is in good agreement with their fluxes listed in the ARTSS1-5 catalog \citep{2024A&A...687A.183S}, which were obtained using a maximum likelihood method and accounting for the telescope's PSF.

The aforementioned confidence limit estimation algorithms were deployed at \url{https://www.srg.cosmos.ru/uplim} maintained by  IKI\footnote{Space Research Institute of the Russian Academy of Sciences, Moscow, Russia}.
Upper limits and confidence intervals on X-ray flux are estimated using data from the {\art} telescope, obtained during four complete sky surveys conducted from December 2019 to December 2021.
The start and end times of the individual surveys are listed in Table \ref{tab:surveys}. 
Count maps in the 4--12 keV band are used as input data. To estimate the count rate, an exposure map corrected for vignetting effects is employed.
For sky data tesselation, we used the HEALPix algorithm \citep{2005ApJ...622..759G}.

\begin{table}[ht]
\caption{List of full {\art} sky surveys used.}
\begin{tabular}{lrrcccll} 
\hline
Survey & Start Date & End Date \\
\hline
ARTSS1 & December 12 2019 & June 11 2020 \\
ARTSS2 & June 11 2020 & December 15 2020 \\
ARTSS3 & December 15 2020 & June 16 2021 \\
ARTSS4 & June 16 2021 & December 19 2021 \\
\hline
\end{tabular}
\label{tab:surveys}
\end{table}

To minimize the influence of background contamination from known X-ray sources on upper limit estimates, regions around the positions of sources listed in the ARTSS1-5 \citep{2024A&A...687A.183S} catalog were excluded. The size of these excluded regions depends on the brightness of each source.
For this purpose, we empirically determined five flux ranges and corresponding exclusion radii (see Table \ref{tab:maskradii}). 
Pixels falling within these excluded regions are not used for photometry and a warning message is displayed. 
When inputting coordinates near the positions of sources listed in the ARTSS1-5 catalog, a relevant list is shown.

\begin{table}[ht]
\centering
\caption{The X-ray flux ranges and the corresponding exclusion radii around sources from the ARTSS1-5 catalog.}
\begin{tabular}{cc}
\hline       
X-ray flux range & Exclusion radius \\
$10^{-10}\ erg\ s^{-1}\ cm^{-2}$&\\
\hline
$<1.25$ & $3.6'$ \\
1.25$-$2.5 & $9'$ \\
2.5$-$20 & $30'$ \\
20$-$200 & $54'$ \\
$>200$ & $2.5\degree$ \\
\hline
\end{tabular}
\label{tab:maskradii}
\end{table}

To ensure high query performance, the count rate and exposure maps are stored in a relational database (DB). 
Consequently, each HEALPix tessellation element corresponds to a record in the DB containing information about counts, exposure, exclusion status, and survey number. 
Upper limit and confidence interval estimates of X-ray flux are possible for both the sum of all surveys and for each survey individually. 
Finally, we note that the {\art} telescope continues to operate in full-sky survey mode, so data will be added as they become available and after the release of future {\art} X-ray catalogs.


\section*{Discussion}
\noindent

\begin{figure}
    \centering
    \includegraphics[width=\linewidth]{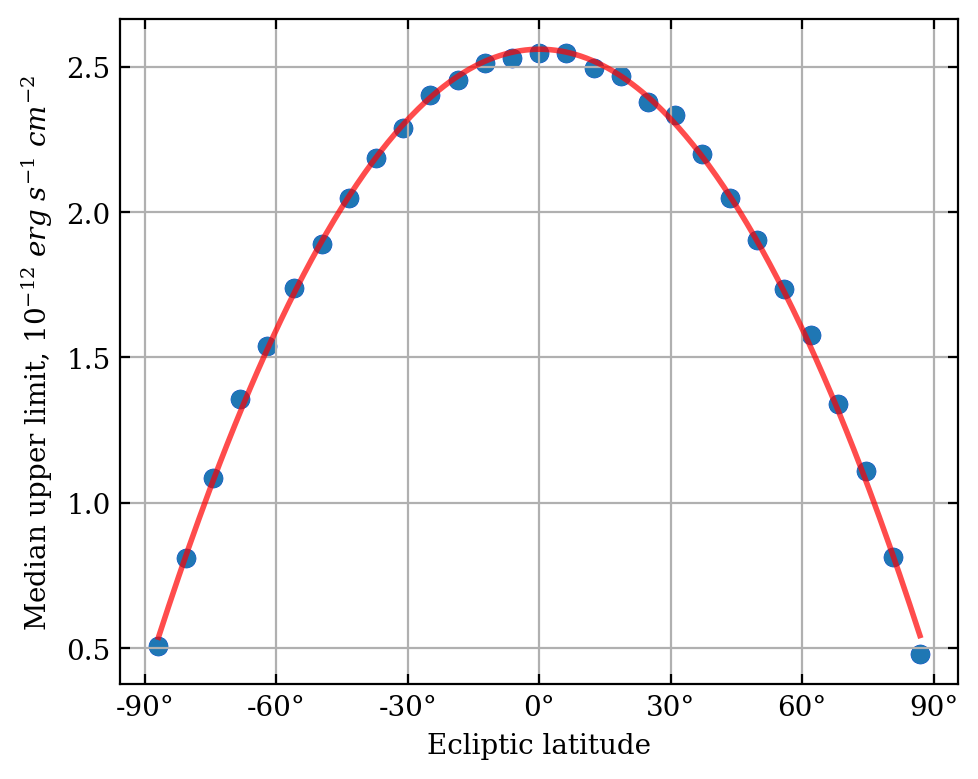}
    \caption{Dependence of the one-sided upper limit (in Bayesian statistics mode) on ecliptic latitude for the sum of the first four surveys, approximated by a second-order polynomial}
    \label{fig:medianupperlimit}
\end{figure}

The SRG observatory's chosen survey strategy results in deeper coverage regions around the north and south ecliptic poles \citep{srg21}, where the great circles of all sky scans intersect.
The minimum and maximum accumulated exposure time occur at the ecliptic equator and poles, respectively. 
The upper limit on X-ray flux in a chosen direction on the sky depends on the exposure -- that is, on the accumulated photon statistics and background particle level -- and is therefore a function of ecliptic latitude.
We performed $10^5$ one-sided upper limit measurements using Bayesian statistics at a 95\% significance level across various sky directions. 
This analysis utilized the combined data from the ARTSS1-4 surveys, excluding regions containing X-ray sources identified in the ARTSS1-5 catalog \citep{2024A&A...687A.183S}.
We then grouped the resulting upper limit values into bins defined by ecliptic latitude intervals, and calculated the median value for each bin.
This analysis reveals a clear dependence of median upper limits on ecliptic latitude, as shown in Fig.~\ref{fig:medianupperlimit}.
This dependence can be approximated with a simple function  $UL=a\theta^2+b\theta+c$, with parameter values being $a=-2.675\cdot10^{-16}\textrm{erg s}^{-1}\text{cm}^{-2}\text{deg}^{-2}$, $b=3.304\cdot10^{-17}\ \textrm{erg s}^{-1}\text{cm}^{-2}\text{deg}^{-1}$, 
$c=2.561\cdot10^{-12}\textrm{erg s}^{-1}\textrm{cm}^{-2}$.

\section{Conclusion}

This work presents a methodology for assessing the sensitivity of X-ray point source detection based on aperture photometry of data from the {\it Mikhail Pavlinsky} {\art} telescope of the SRG observatory.
Both a classical, frequentist approach, and a method relying on Bayes' theorem are considered.
Both one-sided and two-sided confidence limits on the 4-12 keV flux are evaluated for any point on the celestial sphere, using data from the {\art} telescope's full-sky survey.
The software implementation of the methods is available on the SRG project website, utilizing the computational infrastructure of IKI.
To assess the upper limit or confidence interval on the X-ray flux, data from the ART-XC telescope, obtained during the first four full-sky surveys from December 12, 2019, to December 19, 2021, are used.
Using the web interface, users can retrieve these limits using either the combined data from all four surveys or individual year-long survey datasets.
This data can be used to study the X-ray properties of variable and transient objects, including those not detected as significant sources in the ongoing {\art} all-sky survey.

\acknowledgements
The {\it Mikhail Pavlinsky} \art\ telescope is the hard X-ray instrument on board the \srg\ observatory, a flagship astrophysical project of the Russian Federal Space Program realized by the Russian Space Agency in the interests of the Russian Academy of Sciences. The \art\ team thanks the Russian Space Agency, Russian Academy of Sciences, and State Corporation Rosatom for the support of the \srg\ project and \art\ telescope.



\label{lastpage}


\bibliographystyle{astl}
\bibliography{paper}

\end{document}